\documentclass[]{emulateapj}
\usepackage{natbib}
\usepackage{graphicx}
\def\be{\begin{equation}} 
\def\ee{\end{equation}}

\def\msun{{M_\odot}}

\def\HII{\hbox{H~$\scriptstyle\rm II\ $}}

\def\gsim{\lower.5ex\hbox{\gtsima}} 
\def\lsim{\lower.5ex\hbox{\ltsima}} 
\def\gtsima{$\; \buildrel > \over \sim \;$} 
\def\ltsima{$\; \buildrel < \over \sim \;$} 
\def\prosima{$\; \buildrel \propto \over \sim \;$} \def\gsim{\lower.5ex\hbox{\gtsima}} 
\def\lsim{\lower.5ex\hbox{\ltsima}} 
\def\simgt{\lower.5ex\hbox{\gtsima}} 
\def\simlt{\lower.5ex\hbox{\ltsima}} 
\def\simpr{\lower.5ex\hbox{\prosima}}   
  
\def\gtsima{$\; \buildrel > \over \sim \;$} 
\def\ltsima{$\; \buildrel < \over \sim \;$} 
\def\gsim{\lower.5ex\hbox{\gtsima}} 
\def\lsim{\lower.5ex\hbox{\ltsima}} 
\def\simgt{\lower.5ex\hbox{\gtsima}} 
\def\simlt{\lower.5ex\hbox{\ltsima}} 
\def\simpr{\lower.5ex\hbox{\prosima}}

\def\msun{\,{\rm M_\odot}}

\def\E3{{\cal E}_{\rm g}^{III}}

\def\CII{\hbox{C~$\scriptstyle\rm II$}}

\shorttitle{[CII]-SFR relation in high-z galaxies}
\shortauthors{Vallini et al.}

\begin{document}
\title{On the [CII]-SFR relation in high redshift galaxies}

\author{L. Vallini\altaffilmark{1}}
\affil{Dipartimento di Fisica e Astronomia, Universit\'a di Bologna, viale Berti Pichat 6/2, 40127 Bologna, Italy}

\author{S. Gallerani, A. Ferrara\altaffilmark{2}, A. Pallottini, B. Yue}
\affil{Scuola Normale Superiore, Piazza dei Cavalieri 7, 56126, Pisa, Italy}

\altaffiltext{1}{Scuola Normale Superiore, Pisa, Italy}
\altaffiltext{2}{Kavli IPMU (WPI), Todai Institutes for Advanced Study, the University of Tokyo}

\begin{abstract}
After two ALMA observing cycles, only a handful of [$\CII$]
  $158\,\rm{\mu m}$ emission line searches in $z>6$ galaxies have
  reported a positive detection, questioning the applicability of the
local [\CII]-SFR relation to high-$z$ systems. To investigate this
issue we use the \citet[][V13] {vallini2013} model, based on
high-resolution, radiative transfer cosmological simulations to
predict the [$\CII$]  emission from the interstellar medium of a
$z\approx 7$ (halo mass $M_h=1.17\times10^{11}\msun$) galaxy. We
improve the V13 model by including (a) a physically-motivated
metallicity ($Z$) distribution of the gas, (b) the contribution of
Photo-Dissociation Regions (PDRs), (c)  the effects of Cosmic
Microwave Background on the [\CII] line luminosity. We study the
relative contribution of diffuse neutral gas to the total [\CII]
emission ($\rm F _{diff}/F_{tot}$) for different SFR and $Z$
values. We find that the [\CII] emission arises predominantly from
PDRs: regardless of the galaxy properties, $\rm F _{diff}/F_{tot}\leq
10$\% since, at these early epochs, the CMB temperature approaches the
spin temperature of the [\CII] transition in the cold neutral medium
($T_{\rm CMB}\sim T_s^{\rm CNM}\sim 20$~K). Our model predicts a
high-$z$ [\CII]-SFR relation consistent with observations of local
dwarf galaxies ($0.02<Z/Z_{\odot}<0.5$). The [\CII] deficit
  suggested by actual data ($L_{\rm CII}<2.0\times 10^7 L_{\odot}$ in
  BDF3299 at $z\approx7.1$)
  if confirmed by deeper ALMA observations, can be ascribed to
  negative stellar feedback disrupting molecular clouds around star
  formation sites.
The deviation from the local [\CII]-SFR would then imply a modified Kennicutt-Schmidt relation in $z>6$ galaxies. Alternatively/in addition, the deficit might be explained by low gas metallicities ($Z<0.1~Z_{\odot}$).
\end{abstract}

\keywords{galaxies:high-redshift, galaxies:ism, cosmology:theory, submillimeter:ism, line:formation, cosmology:observations}

\section{Introduction}\label{intro}
The study and characterization of the interstellar medium (ISM) of galaxies that formed in the early Universe is entering a golden era thanks to the unprecedented capabilities of the Atacama Large Millimeter-submillimeter Array (ALMA). 
In particular, the $158\,\rm{\mu m}$ emission line due to the $^{2} P_{3/2} \rightarrow ^{2}P_{1/2}$ fine-structure transition of ionized carbon ([\CII]), being the dominant coolant of the neutral
diffuse ISM \citep{wolfire2003}, is by far the brightest line in the far-infrared band
\citep[][]{stacey1991}. 
In addition to the diffuse neutral gas, the [\CII] line can be excited in other components of the interstellar medium such as high
density photodissociation regions (PDRs), 
and in the diffuse
ionized gas, where the main driver of the [\CII] emissivity are
the collisions with free $e^{-}$. 
Although precisely assessing the relative contribution of the various gas
phases to the total line emission might be difficult, [\CII] line
remains an exquisite proxy to characterize the ISM of galaxies that formed during the Epoch of Reionization (EoR; $z\approx6-7$) \citep[e.g][]{carilli2013}.
Before the ALMA advent, the [\CII] line from $z>4$ was solely detected in
galaxies with extreme star formation rates (SFRs)
($\approx$1000~M$_{\odot}\rm yr^{-1}$) \citep[e.g.][]{cox2011,
  carilli2013b, carniani2013, debreuck2014}, or in those hosting
Active Galactic Nuclei (AGN) \citep[e.g.][]{maiolino2005,
  venemans2012, gallerani2012, cicone2015}.
  
In the first years of ALMA operations, the [\CII] emission line has been detected in a handful of galaxies with modest
star formation rates ($50-300\,\rm{M_{\odot}\,yr^{-1}}$) at $z\approx4.5$, i.e. approximately $400\,\rm{Myr}$ after the end of the EoR \citep{carilli2013b, carniani2013, williams2014, riechers2014}. 
Viceversa, other tentative searches of this line have failed in normal star-forming galaxies (NSFGs; SFR $\approx $10~M$_{\sun}\rm yr^{-1}$) at the end of Epoch of Reionization ($z \gtrsim6$) \citep[e.g.][]{walter2012, kanekar2013, gonzalezlopez2014,ouchi2013, ota2014, schaerer2015}. These early results seemed to be at odds with the correlation between the intensity of the
[\CII] line and the SFR observed in local
galaxies \citep{delooze2011, delooze2014}, thus questioning its applicability to sources at $z\gtrsim6$.
Only very recently, three different ALMA campaigns targeting $z\approx5-7$
   Lyman Alpha Emitters (LAEs) and Lyman Break Galaxies (LBGs) have yielded [\CII] detections:
   \citet[][]{maiolino2015}
   in the vicinity of BDF3299, a LAE at $z\approx7.1$,
   \citet[][]{capak2015} in a sample of LAEs at
   $5.1<z<5.7$, and \citet[][]{willott2015} in two
   luminous LBGs at $z\approx6$, being 
  in agreement with the [\CII] luminosity expected from lower-$z$ observations in star forming galaxies.

In the nearby Universe, the [\CII]-SFR relation holds for a wide range of galaxy types,
ranging from metal poor dwarf galaxies, to starbursts, ultra-luminous
infrared galaxies, and AGN hosting galaxies \citep{boselli2002,
  delooze2011, sargsyan2012, delooze2014, pineda2014,
  herrera-camus2015}. 
The [\CII] emission from PDRs is primarily due to the far-ultraviolet
  (FUV) radiation produced by OB stars that form in the vicinity
  of the photodissociation regions \citep{hollenbach1999}. The relation
  between SFR and the [\CII] luminosity in the
  neutral diffuse gas is more subtle.  On the one hand, the [\CII] emissivity is
  proportional to the gas heating due to the photoelectric
  effect on dust grains, namely to the intensity of the FUV radiation \citep{wolfire2003, herrera-camus2015}. On the other
  hand, an increasing FUV radiation reduces the relative abundance of the cold neutral medium (CNM)
  with respect to the warm neutral medium (WNM), thus reducing the [\CII]
luminosity \citep{vallini2013}.
Observational studies have found that, in the plane of the Galaxy, the [\CII] emission is mostly associated with dense PDRs \citep[]{pineda2013}. On the contrary, in low metallicity local dwarf galaxies \citep[e.g. Haro 11,][]{cormier2012}, nearby galaxies \citep[e.g. M51 and M31,][]{kramer2013, parkin2013, kapala2015} and the outskirts of the Milky Way \citep{langer2014} the PDR contribution can be as small as $\approx10\%$. The [\CII]-SFR relation in these cases is shallower than that of starburst galaxies \citep{delooze2014}.

From a theoretical point of view the intensity of the [\CII] line from high-$z$ galaxies has been computed both through numerical simulations \citep[e.g][]{nagamine2006} and semi-analytical models \citep[e.g.][]{gong2012, munoz2014, popping2014}. Recently \citet{olsen2015} present a multi-phased ISM model consisting of molecular clouds, embedded within a cold neutral medium of atomic gas, and hot, partly ionized gas. The model, applied on top of a cosmological SPH simulation of massive star-forming galaxies on the main-sequence at $z=2$, self-consistently calculates the relative contribution of the various phases to the [\CII] emission.

In the previous paper of this series \citet{vallini2013} (hereafter, V13) computed the [\CII] emission arising from the neutral diffuse gas of a single prototypical high-$z$ ($z\approx6.6$) galaxy, extracted from a SPH cosmological simulation, further implemented with radiative transfer calculation. This is crucial to model the intensity of the galaxy internal UV field and the consequent gas ionization structure. The calculation of the [\CII] emission is performed thanks to a sub-grid model describing the thermal equilibrium of the cold and warm neutral medium as a function of the FUV radiation field intensity within the galaxy.
The spatial resolution ($\approx60\,\rm{pc}$) allows to pro\-per\-ly describe the ISM small-scale density structure. 
Here we present an updated version of the V13 model that allows us to also compute the [\CII] emission arising from the clumpy molecular gas, and the effect of the increased CMB temperature on the [\CII] observability. The aim is to finally assess whether the local [\CII]-SFR relation holds at high-$z$, and what we can learn from any deviation from it.

\section{Modelling [\CII] emission}\label{sec:2}
In this Section, we first summarize the main characteristics of the V13 model, referring the interested reader to V13 for further details. Next, we describe the additional features implemented by this work.
\subsection{The V13 model}
We run \texttt{GADGET-2} \citep{springel2005} cosmological SPH
hydrodynamic simulations of a $(10h^{-1}\rm{Mpc})^3$ comoving volume
with a mass resolution of 1.32 (6.68)$\times10^{5}\msun$ for baryons
(dark matter).  We take a snapshot at redshift $z=6.6$, identify
the most massive halo (total mass $M_h=1.17\times10^{11}\msun$,
$r_{vir}\approx20$ kpc), and we select a $(0.625\,h^{-1}\rm{Mpc})^3$
comoving volume around the center of the halo.
We post-process our simulations with the UV radiative transfer (RT) code LICORICE. Gas properties are resolved on a fixed grid with a resolution of $\sim$~60 pc.
We complement the simulation with a sub-grid model taking into account the cooling and heating processes producing a multi-phase thermal ISM
structure \citep{wolfire1995, wolfire2003}. According to this model,
the neutral gas in the ISM is constituted by a two-phase medium in
which the cold neutral medium (CNM) and the warm neutral medium (WNM)
are in pressure equilibrium. The relative abundance of these two
components depends on (i) the gas metallicity, $Z$, determining the
coolants abundance, and (ii) the FUV flux, $G_0$, in
the Habing ($6-13.6\,\rm{eV}$) band, controlling the photoelectric
heating produced by dust grains. The value of $G_0$ scales with
SFR and is calculated as:
\begin{equation}\label{gnot_sfr}
G_0(\vec{r}, \mathrm{SFR})= \mathrm{SFR} \times \Sigma^{n_*}_{i=1}  \frac{ \int_{6\, \mathrm{eV}}^{13.6\,\mathrm{eV}} l_{\nu,i}\rm{d}\nu}{4\pi| \vec{r}-\vec{r_i} |^2},
\end{equation}
where $n_*$ is the number of sources, $\vec{r_i}$ is the positions,
$l_{\nu,i}$ is the monochromatic luminosity per source. We compute
$l_{\nu,i}$ by using \texttt{STARBURST99} template
\citet{leitherer1999}, assuming continuous star formation (SF), an age
$10\,\rm{Myr}$ for the stellar population\footnote{We keep fixed the original assumption of V13 in which a continuous SF with an age of $t_*=10\,\rm{Myr}$ for the stellar population was based on the \emph{Himiko} SED fitting presented in \citet{ouchi2009, ouchi2013}. This scenario have been recently confirmed \citet{zabl2015} and \citet{schaerer2015} that, when adopting continuous/exponential rising (declining) star formation histories for \emph{Himiko}, obtain age of the stellar population in the range $t_*=1-35\,\rm{Myr}$ \citep{zabl2015}, and $t_*=10-40\,\rm{Myr}$ \citep{schaerer2015}}  and setting the metallicity
accordingly to the cell value (see Sec.\ref{sec:metallicity}). In this work we explore the range
SFR$=[0.1-100] \,\rm M_{\odot}\,yr^{-1}$. For each value of
  the star formation rate and metallicity we run the V13 sub-grid
  model computing the expected distribution of the cold and warm
  diffuse gas within the galaxy.

We calculate the [\CII] emissivity through Eq. (3) of V13. We note that this equation is valid under the assumption that the density of the colliding species (electrons and hydrogen atoms) is much lower than their corresponding critical densities\footnote{The critical density for collision with neutral hydrogen atoms (or with $e^{-}$) has been computed at $T=100$ K, a value consistent with the CNM temperature.} and that no external radiation field is present. While the first assumption is justified by the fact that in our simulation $n_e\ll n^e_{crit}=8\,\rm{cm^{-3}}$ and $n_H\ll 3000 \,\rm{cm^{-3}}$, in Sec. 2.3 we investigate whether the CMB may affect the intensity of the [\CII] emission.
As pointed out in V13, the CNM accounts for $\approx 95\%$ of the total [\CII] emission arising from the diffuse neutral
medium. Given this result, we refer to the diffuse medium as CNM.
 
\subsection{Metallicity}\label{sec:metallicity}
The V13 model assumes that metals are uniformly distributed within the galaxy. Hereafter, we refer to models with a uniform metallicity distribution as\footnote{For example, we indicate with C02 a model in which a uniform $Z=0.2~Z_{\odot}$ is imposed.} ``C-models''. In this work, we also consider the possibility that the distribution of metals follows the density distribution, by relating $Z$ to the baryonic overdensity of each cell $\Delta \equiv \rho_{gas}/\rho_{c}(z)$, where $\rho_{gas}$ is the gas density in the cell and $\rho_{c}(z)$ is the critical density at redshift $z$.
This is in agreement with the parametrization adopted by \citet{keating2014} to describe the circumgalactic medium of high-$z$ galaxies, and with the results presented in \citet[][P14]{pallottini2014} that has been applied to our galaxy.

\begin{figure}
\centering
\includegraphics[scale=0.33]{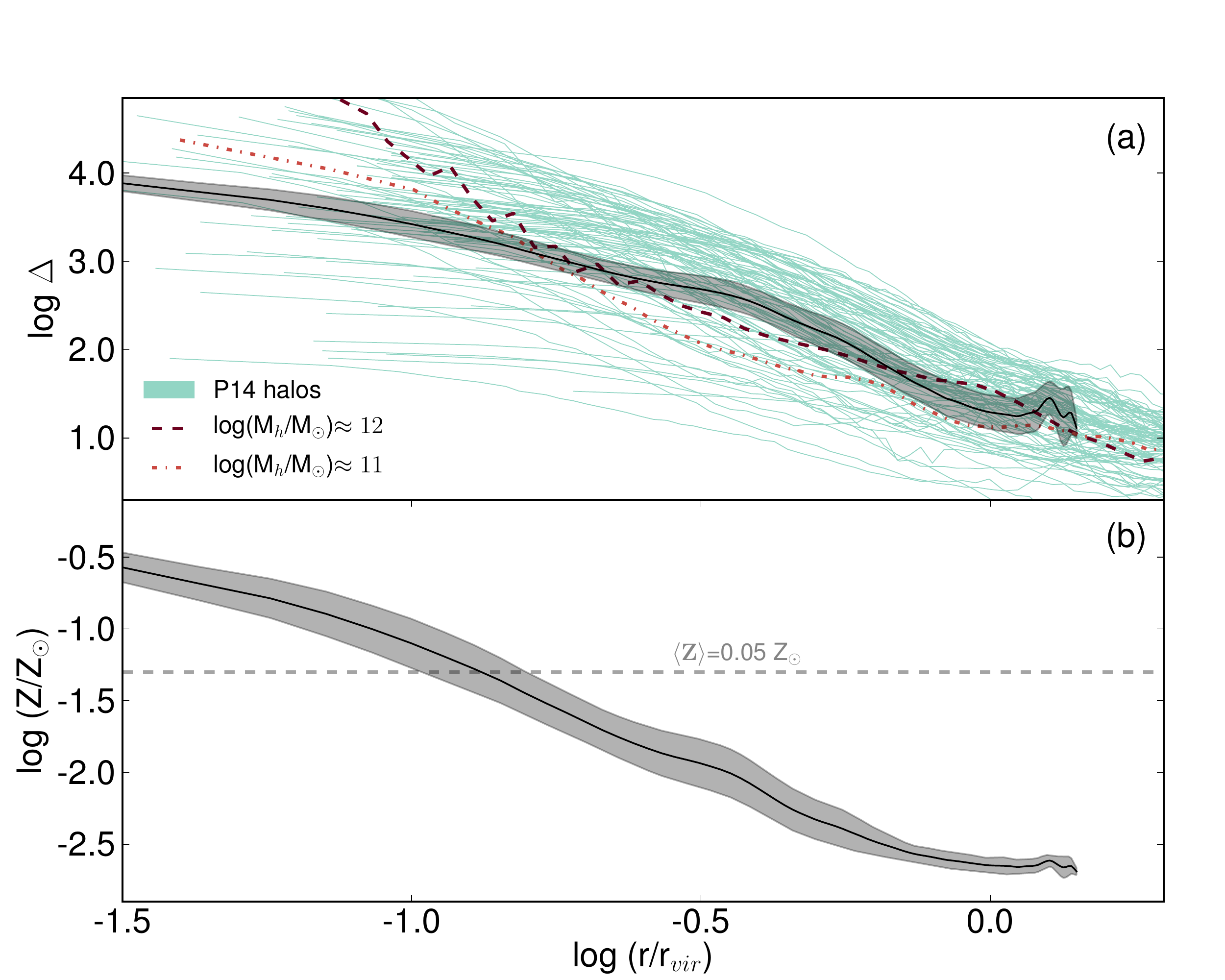}
\caption{\emph{Panel (a)}: The black solid line and the gray shaded region represent the radial profile of the baryonic overdensity ($\Delta$) and
its r.m.s. fluctuation for our simulated galaxy. The distance is rescaled with the virial radius of the galaxy. With light cyan lines we plot the mean
profiles for the $\sim100$ galaxies in the \citet[][(P14)]{pallottini2014} simulation (see also Fig. 1 in \citep{pallottini:2014_cgmh}). We highlight with
solid dashed/dot-dashed lines two most massive P14 galaxies. Such galaxy are hosted in dark matter halos with mass $\log(M_h/M_{\sun})\approx12$ and $\log(M_h/M_{\sun})\approx11$ respectively, comparable the one of the galaxy adopted in this work ($\log(M_h/M_{\sun})\approx11.1$). \emph{Panel (b):} Metallicity ($Z$) radial profile for the P005' model. For the P005, $Z$ is calculated using the density of our simulated galaxy and rescaling $Z-\Delta$ relation found in P14. The relation is rescaled so the mean metallicity of the galaxy is $\langle Z\rangle=0.05\,\rm{Z_{\odot}}$ (grey dashed line). Note that $\langle Z\rangle$ is calculated considering only those cells whose baryonic overdensity is $\Delta > 200$, i.e. up to $r/r_{vir}\approx0.5$. The MCs are, on average, clusterized near the center, and this affects their mean metallicity.}\label{profile}
\end{figure}

P14 use a customized version of the adaptive mesh refinement code RAMSES \citep{teyssier:2002} in order to investigate the metal enrichment of high-$z$ galaxies. In P14 star formation is included via subgrid prescriptions, and supernovae feedback is accounted by implementing a metal-dependent parameterization of stellar yields and return fractions based on population synthesized models. The P14 galaxy sample reproduces the observed cosmic star formation rate \citep[][]{Bouwens:2012ApJ,Zheng:2012Natur} and stellar mass densities \citep[][]{Gonzalez:2011} evolution in the redshift range $4\leq z \lsim 10$. In the upper panel of Fig. \ref{profile} we plot the radial profile of the baryonic overdensity ($\Delta$) and its r.m.s. fluctuation of our simulated galaxy. We test the profile with a sample of $\sim100$ galaxies extracted from P14 (see also \citealt{pallottini:2014_cgmh}, in particular Fig. 1, upper panel). Among the P14 galaxies, with solid dashed/dot-dashed lines we highlight two galaxy hosted in dark matter halo with mass $\log(M_h/M_{\odot})\approx12$ and $\log(M_h/M_{\odot})\approx11$, respectively. This is comparable to the dark matter halo mass of the galaxy adopted in this work ($\log(M_h/M_{\odot})\approx11.1$).

In particular, P14 found a tight correlation between $Z$ and $\Delta$ for $\log\Delta\gsim 2$, namely for overdensities typical of galaxy outskirts/ISM.
While in the IGM ($\log\,\Delta \leq 2.3$) the metallicity is only weakly correlated with $\Delta$, in the ISM ($ 2.3 < \log \,\Delta < 4.5$) the $Z-\Delta$ relation is tight. This is due to the fact that the most overdense regions denote the location in which stars form, and that are therefore more efficiently polluted with metals.
We fit the $Z-\Delta$ relation provided in their paper and we
normalize the relation to the mean metallicity $\langle Z \rangle$
over the galaxy, i.e. over those cells whose overdensity is $\Delta >
200 $. Hereafter, we refer to models that take into account this $Z-\Delta$ relation as ``P-models''. The density-dependent metallicity case with $\langle Z \rangle=0.05 Z_{\odot}$ is called P005, and the profile is shown in the (b) panel of Fig. \ref{profile}.

\subsection{Molecular clouds and PDRs}\label{including_pdr}
Beside the emission arising from the diffuse neutral medium, [\CII] line can be excited in the so-called photodissociation regions \citep{tielens1985, hollenbach1999} around molecular clouds (MCs).
To establish whether the gas in a cell becomes gravitationally bound, we apply the Jeans instability criterium.
We define \emph{molecular cells} those satisfying the following condition: $M_{cell} > M_{J}(T_{\rm CMB}, n_{cell})$, where $M_{cell}$ is the mass of the gas in the cell, $M_{J}$ is the Jeans mass at the CMB temperature ($T_{\rm CMB}$) at $z=6.6$, and at the density of the gas in the simulation cell ($n_{cell}$). As a caveat we note that is possible that other processes such as the cosmic rays heating \citep[e.g.][]{papadopoulos2010}
can increase the temperature above the CMB floor, even at $z\approx6$. By applying this prescription, we find that the total mass of molecular hydrogen in the simulated galaxy is $M_{\rm{H_2}}=3.9\times10^{8}\rm{M_{\odot}}$, consistently with previous theoretical estimates \citep[$M_{\rm H_2}= 2.5\times 10^{8}\,\rm M_{\odot}$,][]{vallini2012} and observational constraints \citep[$M_{\rm H_2}<4.9\times10^{9}\,\rm M_{\odot}$,][]{wagg2009} on $z\approx7$ LAEs.

Since we find that on average the simulation cells contain a molecular hydrogen mass $\langle M^{cell}_{\rm H_2} \rangle \sim 3 \times 10^{3}\,\rm{M_{\odot}}$, we consider each molecular cell as a Giant Molecular Cloud \citep[GMC; $M_{\rm GMC}=10^3-10^6\,\rm{M_{\odot}}$,][]{murray2011}.
The properties of GMCs are controlled by a turbulent and highly supersonic velocity field that causes isothermal shock waves \citep{padoan1995, ostriker2001, padoan2014}.
The problem of turbulent fragmentation of molecular clouds can be treated analytically \citep[e.g.][]{krumholz2005, padoan2011, hennebelle2011, hennebelle2013} or numerically \citep[e.g.][]{vazquez-semadeni1994, kim2002, kim2003, wada2008, tasker2009}.
Analytical models as well as numerical simulations show that the distribution of the gas density ($n_{cl}$) in an isothermal, non self-gravitating, turbulent medium follows a log-normal distribution \citep{padoan1995, padoan2011}:
\begin{equation}\label{padoan_pdf}
p(\ln x) d\ln x = \frac{1}{(2 \pi \sigma^2)^{1/2}} {\rm exp}\left[ -\frac{1}{2} \left( \frac{{\ln\,} x - \overline{{\ln\,} x}}{\sigma} \right)^2 \right],
\end{equation}
where $x=n_{cl}/\overline{n_{cl}}$, $\overline{n}_{cl}\simeq n_0\,\mathcal{M}^2$, $n_0$ is the average number density of the CNM ($n_0=50 \,\rm{cm^{-3}}$, see V13), $\mathcal{M}=10$ is the Mach number value suggested by \citet{kainulainen2013}, the turbulent velocity dispersion is given by $\sigma=\sqrt{\rm{ln}[1+(\mathcal{M}\beta^{2})]}$ with $\beta=0.5$ and $\overline{\ln x}=-0.5\sigma^2$.
If we assume that each GMC in our simulation is composed by a set of clumps, we can compute the densities $n_{cl}$ of each clump by adopting an iterative approach that consists of the following steps: 
\begin{enumerate}
\item[1.] Select $n_{cl}$ by sampling the density distribution (Eq. \ref{padoan_pdf}).
\item[2.] Set the clump radius equal to the Jeans length $r_{cl}=\lambda_J(T_{CMB}, n_{cl})$.
\item[3.] Calculate the clump mass $M_{cl} = (8\pi/3) m_H n_{cl} r_{cl}^3 $.
\item[4.] If $M_{cl}<M_{cell}$ calculate the residual mass in the cell $M_{cel}^{new}=M_{cell}-M_{cl}$; go to step 1.
\item[5.] If $M_{cl}>M_{cell}$ reject the density sampled and assume $M_{cl}=M_{cell}$. Calculate $r_{cl}$ as in step 2. and use it to derive the resulting clump density.
\end{enumerate}
Through this procedure, we find that the MCs in the simulations are
characterized by the following properties: $\langle \log (n_{cl}/
cm^{-3})\rangle  = 2.9 \pm 0.4$, $\langle M_{cl}\rangle = 50\pm 20\,\rm{M_{\odot}}$ and radius $\langle r_{cl} \rangle = 0.7\pm 0.3\,\rm{pc}$.

\begin{figure}
\includegraphics[scale=0.35]{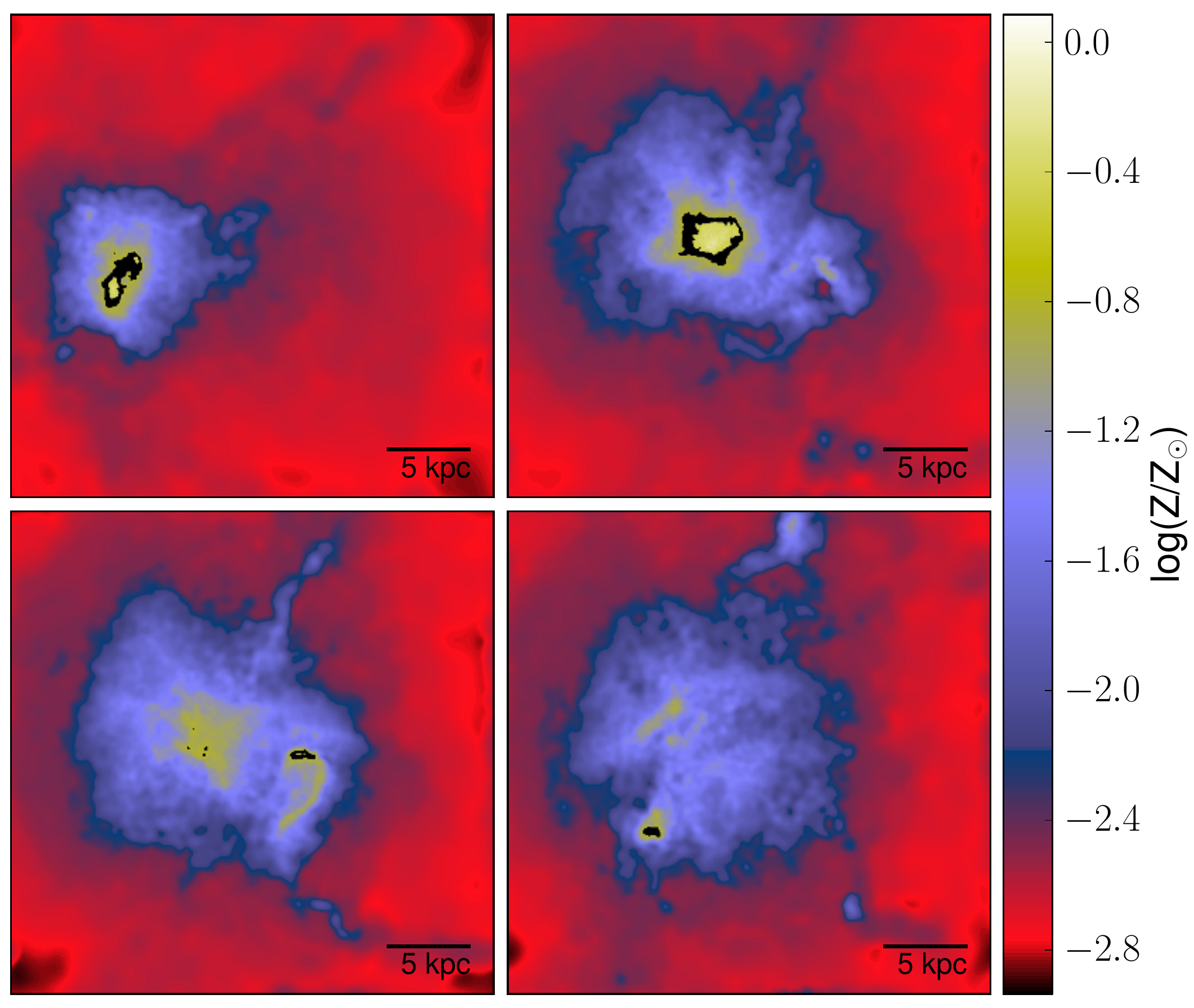}
\centering
\caption{Slices of thickness 57 pc cut through the simulated galaxy
  for the P005 model showing the metallicity distribution (color
    scale) and molecular clouds (black regions). Note that the metallicity of the overdense regions in which MCs reside is greater than the mean value $\langle Z \rangle = 0.05\,\rm{Z_{\odot}}$.}\label{metal_stars_clouds}
\end{figure}

In Fig. \ref{metal_stars_clouds}, we show four 57 pc-thick slices of
the simulated box showing the metallicity distribution and molecular
clouds (black regions) for the P005 model. 
By inspecting the maps shown in the Figure we note that
molecular cells reside in highly overdense regions mainly clustered at the center of the galaxy.
The molecular cells are located predominantly in the innermost region
of the galaxy ($d_{MCs}\approx 1\,\rm{kpc}$) and have $Z \approx 0.2\,\rm{Z_{\odot}}$, whereas the mean metallicity over the galaxy is
$\langle Z \rangle = 0.05\,\rm{Z_{\odot}}$.
This implies that any line arising from the PDRs would provide an upper limit on the mean metallicity of the galaxy.
Finally, to calculate the [\CII] emission from PDRs,  we couple our simulation with \texttt{UCL\_PDR} \citep{bell2005, bell2007, bayet2009} a PDR code that allows to derive the [\CII] emissivity as a function of the intensity of the FUV radiation field ($G_0$, Sec. 2.1), metallicity ($Z$, Sec. 2.2) and molecular gas density ($n_{cl}$, Sec 2.3). 

\subsection{CMB effects on [\CII] emission}\label{CMB_eff}
Since the CMB sets the minimum temperature of the ISM to $T_{\rm
  CMB}=T^0_{\rm CMB}(1+z)$, at high $z$ it represents a strong
background against which the line fluxes are detected
\citep[e.g.][]{dacunha2013}. The contrast of the cloud emission against the CMB radiation is given by the following relation:
\begin{equation}
\Delta I_{\nu} = \left[ B_{\nu}(T_{s}) - B_{\nu}(T_{\rm CMB})\right] (1-e^{-\tau_{\nu}}),
\end{equation}
where $T_s$ is the excitation (or spin) temperature. If we assume that the [\CII] line is optically thin in the sub-millimeter, i.e. $e^{-\tau_{\nu}}\approx 1- \tau_{\nu}$, the ratio ($(1+z)F_{\nu} \sim I_{\nu}/d_L^{2})$) between the flux observed against the CMB and the intrinsic flux emitted by the cloud is \citep[see also][]{dacunha2013}:
\begin{equation}\label{zeta}
\zeta \equiv \frac{F_{\nu}^{\rm ag}}{F_{\nu}^{\rm int}}=\frac{ \left[ B_{\nu}(T_{s}) - B_{\nu}(T_{\rm CMB})\right] \tau_{\nu}}{B_{\nu}(T_{s})\tau_{\nu}}= 1- \frac{B_{\nu}(T_{\rm CMB})}{B_{\nu}(T_{s})}.
\end{equation}
Eq. (\ref{zeta}) clearly shows that as the $T_{\rm CMB}$ approaches $T_{s}$ the observed flux tends to zero.
The ratio of the population of the upper ($^2P_{3/2}$, labeled
  with $u$) and lower ($^2P_{1/2}$, labeled with $l$) levels of the
  [\CII] 158 $\rm\mu m$ transition formally defines the spin temperature:
\begin{equation}\label{level_pop}
\frac{n_u}{n_l} = \frac{B_{lu}I_{\nu} + n_e C^e_{lu} + n_H C^H_{lu}}{B_{ul}I_{\nu} + A_{ul} + n_e C^e_{ul} + n_H C^H_{ul}} \equiv \frac{g_u}{g_l} e^{-T_{*}/T_{\rm s}},
\end{equation}
where $A_{ul}$ is the Einstein coefficient for spontaneous emission, $B_{ul}$ ($B_{lu}$) is the stimulated emission (absorption) coefficient, $C^{e}_{lu}$ ($C^{H}_{lu}$) is the collisional excitation rate for collision with $e^-$ (protons),  $C^{e}_{ul}$ ($C^{H}_{ul}$) is the collisional de-excitation rate for collision with $e^-$ (protons), and $n_e$ ($n_H$) is the number density of $e^-$ (protons). For the [\CII] $158\,\rm{\mu m}$ line emission$A_{ul}=2.36 \times 10^{-6}\,\rm{s^{-1}}$, $C^e_{lu}(T)= (8.63 \times 10^{-6}/g_l \sqrt{T})\gamma_{lu}(T) e^{-T_*/T}$, with $\gamma_{lu}(T)\approx 1.6$ if $100 < T < 10^3$ \citep{gong2012}, $C^{H}_{lu}(T)$ is tabulated in \citet{dalgarno1972}, and $T_*\equiv h \nu_{ul}/k_b\approx 91 \,\rm{K}$.

In local thermal equilibrium (LTE) the collisional excitation and de-excitation rates are related by the following expression that depends on the kinetic temperature $T$:
\begin{equation}\label{boltz}
C^{e,H}_{lu}=\frac{g_u}{g_l} e^{-(T_*/T)} C^{e,H}_{ul}.
\end{equation}
By combining the equations (\ref{level_pop}) and (\ref{boltz}) we obtain:
\begin{equation}
\label{tspinequation}
\frac{T_*}{T_{\rm s}} = ln \frac{A_{ul}(1+\frac{c^2 I_{\nu}}{2h\nu ^3}) + n_e C^e_{ul} + n_H C^H_{ul}}{A_{ul}(\frac{c^2 I_{\nu}}{2h \nu^3}) + n_e C^e_{ul} e^{-\frac{T_*}{T}} + n_H C^H_{ul} e^{-\frac{T_*}{T}}}.
\end{equation}
As discussed by \citet{gong2012}, the soft UV
background at 1330 \AA~ ($I_{\rm UV}$) produced by stars can in
principle pump the [\CII] ions from the energy level $^2P_{1/2}$ to
$^2D_{3/2}$. This pumping effect, can lead to the [\CII] fine
structure transition $^2D_{3/2}\rightarrow ^2P_{3/2} \rightarrow
^2P_{1/2}$, which would mix the levels of the [\CII] 158 $\rm\mu m$
line and thus modify the Eq. (\ref{tspinequation}). However, the UV pumping effects are negligible in our calculations since the UV intensity inside the galaxy for all the SFR values considered is much smaller than the critical value for this effect to become important, namely $10^{-15}$erg~s$^{-1}$~cm$^{-2}$Hz$^{-1}$~sr$^{-1}$ \citep{gong2012}.\\
We calculate the spin temperature of the [\CII] transition in the PDRs by substituting into  Eq. (\ref{tspinequation}) $n_{e},\,
  n_{H}$ and $T$ as resulting from the \texttt{UCL\_PDR} outputs. The gas temperature within PDRs depends on the radius
  considered and on the SFR and ranges between $20.7<\langle T\rangle
  < 800 \,\rm{K}$.
We find that $T_{\rm s}^{PDR}\sim 30-120\,K$ for SFR$=0.1 -100\,M_{\odot}\,\mathrm{yr}^{-1}$.
The $T_s$ in the CNM is calculated by considering the $n_e$, $n_H$, and $T_k$ provided by the V13 sub-grid model. We obtain $T_{\rm s}^{CNM}\sim 22-23$~K approximately constant in the range of SFR considered.\\
Since at $z\approx6.6$ $T_{\rm CMB}\approx 20.7\,\rm{K}$, we find that the [\CII] emission arising from PDRs is only slightly affected by the CMB ($\zeta \approx 0.8 - 1.0$). Viceversa, the CNM is strongly attenuated at this redshift ($\zeta \approx 0.1 - 0.2$); in this case the CMB effect becomes negligible only for galaxies at $z\leq 4.5$.

\section{Results}
\begin{figure}
\centering
\includegraphics[scale=0.4]{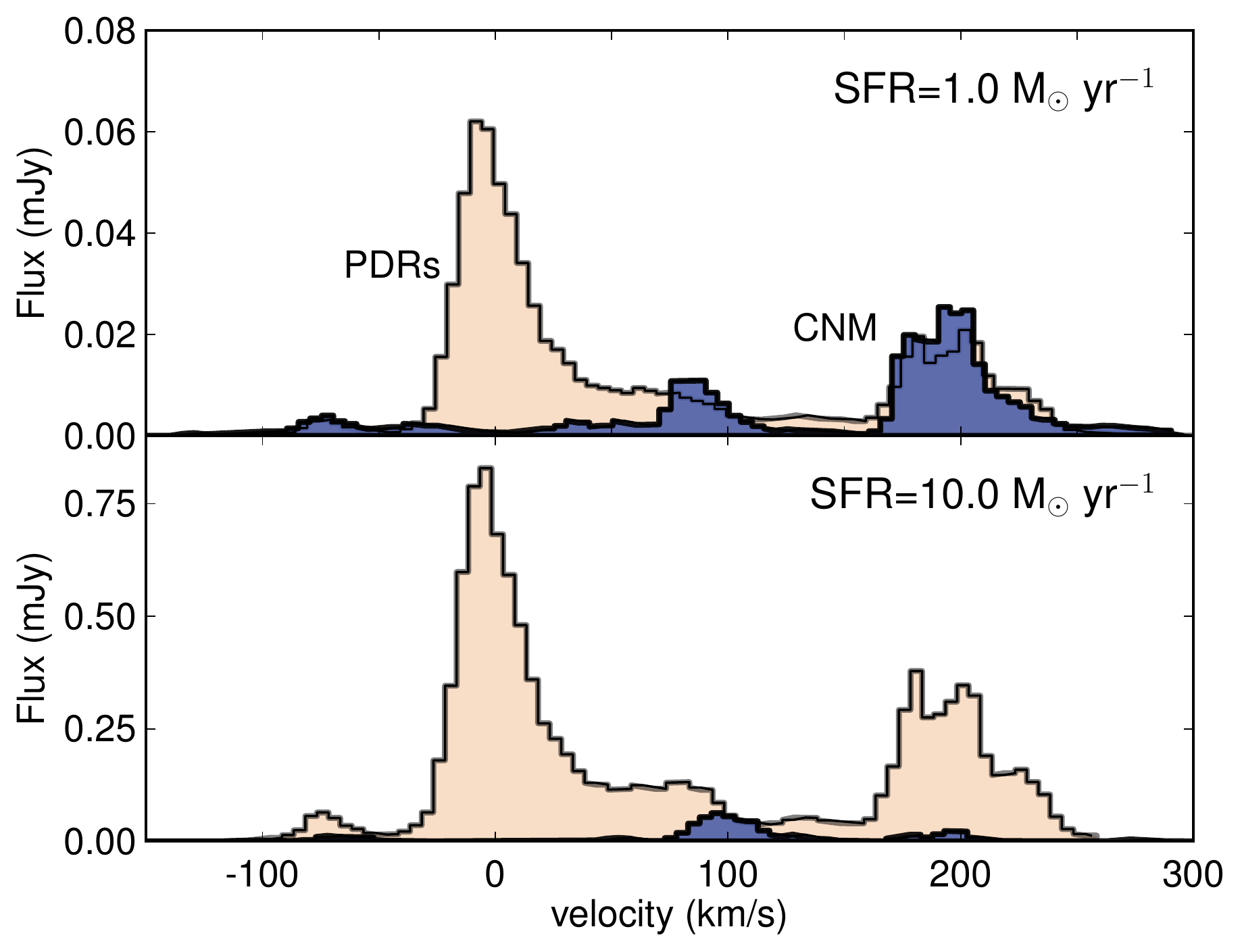}
\caption{[\CII] spectrum for the P005 model assuming $SFR=1\,\rm{M_{\odot}\,yr^{-1}}$ (upper panel) and SFR=$10\,\rm{M_{\odot}\,yr^{-1}}$ (lower panel) and rebinned over $5\,\rm{km\,s^{-1}}$ velocity channels. The emission from PDRs (diffuse neutral medium) is plotted in light red (dark blue).}\label{CII_spectrum}
\end{figure}

\begin{figure}
\centering
\includegraphics[scale=0.38]{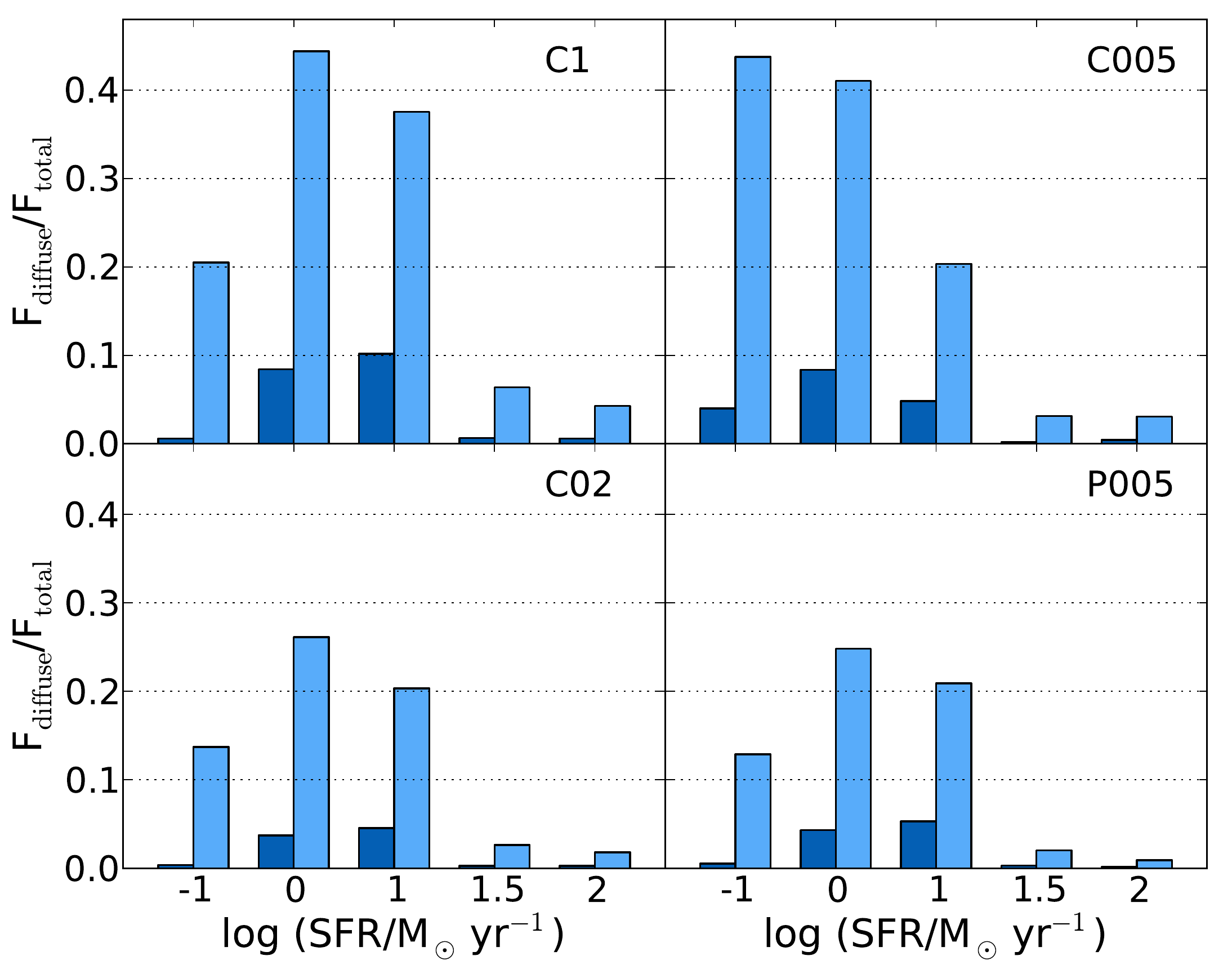}
\caption{Fraction of the [\CII] flux arising from the diffuse medium ($F_{\rm diff}$) over the total flux ($F_{\rm tot}$), for four different models, as a function of the SFR, with (dark blue) and without (light blue) taking into account CMB effects on the [\CII] emission.}\label{fractions}
\end{figure}

In Fig. \ref{CII_spectrum}, we show the [\CII] spectrum obtained from the P005 model, assuming SFR = $1\,\rm{M_{\odot}\,yr^{-1}}$ (top panel) and SFR = $10\,\rm{M_{\odot}\,yr^{-1}}$ (bottom panel). In this Figure, the contribution to the [\CII] emission arising from PDRs and the CNM is shown in light red and dark blue, respectively. 
The emission from PDRs arises predominantly from the center of the galaxy, covering the velocity channels around $\sim0\,\rm{km\,s^{-1}}$.
The second peak in the PDR emission at $v \sim 200\,\rm{km\,s^{-1}}$ is produced by MCs located in the CNM clumps at the periphery of the galaxy (see Fig. \ref{metal_stars_clouds}). 
[$\CII$] emission from the diffuse medium, visible as the two peaks around $\sim 100\,\rm{km\,s^{-1}}$ and $\sim 200\,\rm{km\,s^{-1}}$ is instead always displaced from the center of the galaxy. 
The [\CII] line is relatively narrow, with a FWHM$\sim50\,\rm{km\,s^{-1}}$, as in V13.
In Fig. \ref{fractions}, we plot the relative contribution of the diffuse medium to the total [\CII] emission,  $\rm F_{diff}/F_{tot}$, as a function of the SFR, for different C- and P-models, taking into account the CMB attenuation (dark blue). We find that the [\CII] emission in $z\approx 6$ galaxies is dominated by PDRs, since the CNM contribution is always $\leq 10\%$, regardless of the metallicity profile and SFR considered.

When the CMB attenuation of the CNM luminosity is negligible (i.e. typically for sources located at  $z\leq 4.5$ see Sec. \ref{CMB_eff}), we find $\rm F_{diff}/F_{tot}= 0.05 - 0.45$, consistently with several observations of [\CII] emission in nearby galaxies \citep{cormier2012, kramer2013, parkin2013, langer2014}. 

\subsection{The [\CII]-SFR relation}

In the previous Section, we have found that the [\CII] emission is dominated by PDRs, implying $L_{\rm CII}\propto M_{\rm H_2}$.  We rescale the  [\CII] luminosity of our ``fiducial'' model (SFR$=10\,\rm{M_{\odot}\,yr^{-1}}$, $M_{\rm H_2} = 4 \times 10^{8} M_\odot$) to an arbitrary molecular content by assuming the Kennicutt-Schmidt (KS) relation \citep{kennicutt1998, kennicutt2012}, namely a power-law correlation between the SFR and molecular gas surface densities, $\Sigma_{\rm SFR} \propto \Sigma_{\rm H_2}^N$. 
The range in power-law index ($N$) relating $\Sigma_{SFR}$ and $\Sigma_{\rm H_2}$ depends on a variety of factors, among which the most important ones are the observed scale, and the calibration of star formation rates. \citet{kennicutt1998} and \citet{narayanan2012} report super-linear indeces $N =1.4$ and $N\approx2$ respectively, while \citet{bigiel2008} inferred an approximately linear molecular KS relation. More recently \citet{shetty2013}, by performing a hierarchical Bayesian analysis on the same sample considered by \citet{bigiel2008} conclude that $N=0.84$, with $2\sigma$ range [0.63--1.0]. In what follows we scale the molecular mass with the SFR by adopting $N=1$ leaving to the last Section the discussion on the impact of different $N$ on [\CII]-SFR relation.

\begin{figure*}
\centering
\includegraphics[scale=0.4]{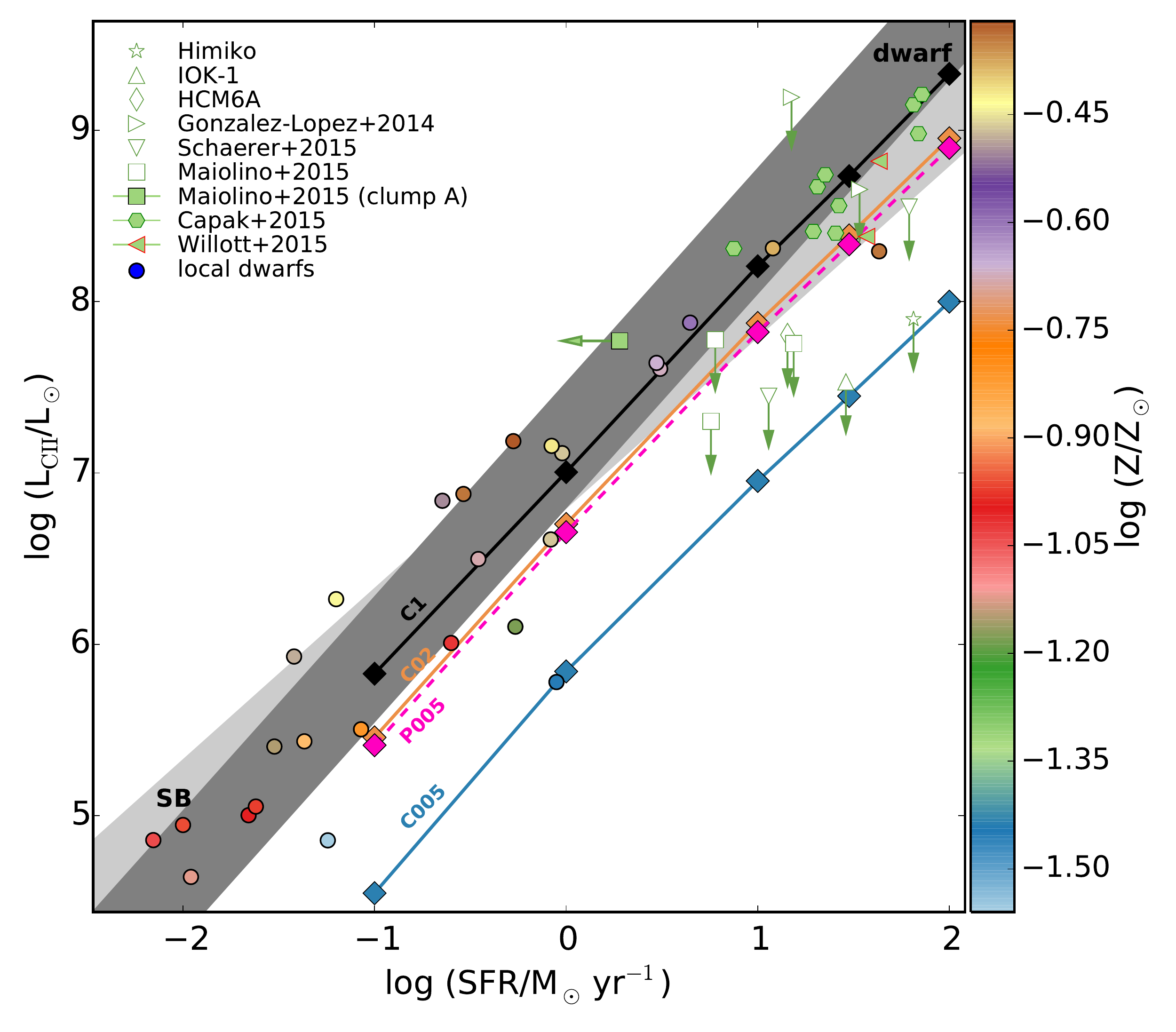}
\caption{[\CII] luminosities in solar units as a function of the SFR. Results from this work are shown with big diamonds and lines color coded as a function of the metallicity. Solid lines represent the result obtained assuming a constant metallicity: black for $Z=Z_{\odot}$ (C1), orange for $Z=0.2\,\rm{Z_{\odot}}$ (C02), and blue for $Z=0.05\,\rm{Z_{\odot}}$ (C005). The magenta dashed line corresponds to a density-dependent
  metallicity (see Sec. 2.2) with $\langle Z \rangle =0.05\,Z_{\odot}$
  (P005). Model predictions are compared
  with data of local dwarf galaxies individually denoted by circles
    color-coded according to their metallicity \citep{delooze2014}. The
    $1\sigma$ scatter around the best fit relation for dwarf and local starburst galaxies
    is plotted in dark gray and light gray, respectively. Green empty (filled) points represent upper limits
    (detections) of [\CII] in LAEs and LBGs at $z\approx5-7$. More
    precisely: the filled square represents the [\CII] detection nearby BDF3299 reported by \citet{maiolino2015}, filled hexagons show recent observations
    by \citet{capak2015}, filled triangles with red
    edges denote data by \citet{willott2015}. Upper
    limits on BDF3299, BDF512, and SDF46 \citep[][]{maiolino2015} are plotted with empty
    squares, along with the upper limits on Himiko
    \citep{ouchi2013, ota2014} (empty star), IOK-1 \citep{ota2014}
    (up-triangle), HCM6A \citep{kanekar2013} (empty diamond), A1703-zD1
    and z8-GND-5296 \citep{schaerer2015} (empty down-triangle), SDF1543 and SDF3058
    \citep{gonzalezlopez2014} (empty left triangle).}\label{CII_sfr_relation}
\end{figure*}
In Fig. \ref{CII_sfr_relation}, we show the result of this procedure for different metallicity profiles. Models with uniform metallicity are shown with a black dotted line for $Z=Z_{\odot}$ (C1), orange solid line for $Z=0.2\,\rm{Z_{\odot}}$ (C02), and blue dotted-dashed line for $Z=0.05\,\rm{Z_{\odot}}$ (C005). The results from our C-models are well described by the following best-fitting formula:
\begin{eqnarray}
{\rm log}{L_{\rm CII}}=7.0+1.2~{\log} (\mathrm{SFR}) +0.021~{\log} (Z) +\nonumber\\
0.012~{\log (\mathrm{SFR})\log (Z)}-0.74~{\log ^2(Z)},
\end{eqnarray} 
where $\rm L_{CII}$ is expressed in solar units, and the SFR in $\rm M_{\odot}\,yr^{-1}$.
The magenta dashed line indicates the predictions for the $Z$-$\Delta$
relation with $\langle Z\rangle = 0.05 Z_{\odot}$ (P005). The slope of
the [\CII]-SFR relation does not depend neither on $\langle Z\rangle$
nor on the metallicity distribution. Moreover, the $[\CII]$ luminosity
predicted by the P005 model is almost coincident with that obtained
from the C02 model, in the entire range of SFR  considered. This can
be understood by considering that, in the case of a overdensity-dependent metallicity profile the photodissociation regions located ad the center and dominating the emission, have $Z_{\rm PDR}>\langle Z \rangle$ and, more precisely, $Z_{\rm PDR}\approx 0.2 \, \rm{Z_{\odot}}$, namely the metallicity value of the C02 model. 

We compare our predictions with the [\CII]-SFR relations and their 1$\sigma$ scatter found by \citet{delooze2014} for local dwarf galaxies (dark gray shaded region) and local starburst galaxies (light gray shaded region). Dwarf galaxy data by \citet{delooze2014} are shown through filled circles, individually color-coded according to their $Z$. Upper limits on the [\CII] luminosity from $z\approx 6-7$ LAEs and LBGs data are indicated with empty symbols \citep{ouchi2013, ota2014, gonzalezlopez2014, schaerer2015, maiolino2015}. The recent [\CII] detection in the vicinity of a $z\approx 7.1$ LAE \citep{maiolino2015} is plotted as a filled square, while [\CII] data at $z\approx 5-6$ by \citet{capak2015} and \citet{willott2015} are indicated with filled hexagons and triangles, respectively.
The [\CII]-SFR relation predicted by our model fairly reproduces the slope of the relation found in local dwarfs, as well as its trend with metallicity, although the scatter in the data is large. 
As shown in Fig. 5, the [CII] emission arising from the diffuse medium is expected to be $\leq 40$\% in local galaxies, where the CMB attenuation on the CNM luminosity is negligible. This implies that the [CII]-SFR relation is always driven by the correlation between the SFR and the intensity of  the [CII] emission arising from PDRs, with and without taking into account CMB effects. Possible variations in the diffuse medium contribution to the total [CII] emission may result in a slight tilt of the [CII]-SFR relation, certainly within the current 1$\sigma$ scatter.

\section{Summary and Discussion}\label{sec:5}
By coupling radiative transfer cosmological simulations of a $z=6.6$ galaxy with a sub-grid ISM model and a PDR code (\texttt{UCL\_PDR}), we have computed the [\CII] emission arising from the diffuse cold neutral medium (CNM) and molecular clouds in early galaxies, characterized by SFRs ranging from 0.1 to 100 M$_{\odot}$yr$^{-1}$. We have distributed metals in the ISM, both uniformly and according to the $Z-\Delta$ relation found by \citet{pallottini2014}, to simulate gas metallicities in the range 0.05-1 $Z_{\odot}$.

We find that the [\CII] line from high-$z$ galaxies is dominated by emission from PDRs, while the CNM accounts for $\leq 10$\% of the total flux. This is due to the fact that at these early epochs the CMB temperature approaches the spin temperature of the [\CII] transition in the CNM ($T_{\rm CMB}\sim T_s^{\rm CNM}\sim 20$~K) suppressing the flux contrast.  The [\CII] spectrum predicted by our model is complex. It shows a pronounced peak (FWHM$\sim 50$~km~s$^{-1}$) due to centrally located ($v=0$) PDRs, and weaker [\CII] displaced  ($v\sim 200$~km~s$^{-1}$) peaks from MCs in the galaxy outskirts. 

The predicted [\CII]-SFR relation reproduces the corresponding relation found in local dwarfs remarkably well. Current upper limits from observations of $z\sim 6-7$ LAEs and LBGs seem to indicate that these galaxies are characterized by a [\CII] luminosity fainter than expected from the local relation. Although this conclusion is still not definitive, it must be noted that the SFRs quoted for high-$z$ galaxies are inferred from observations of the Ly$\alpha$ emission line, and therefore must be considered as lower limits to the actual value. This implies that green arrows in Fig. \ref{CII_sfr_relation} should be moved towards higher SFR values, hence exhacerbating the inconsistency with the local relation.
\begin{deluxetable*}{llllllllll}
\tablecolumns{10}
\tablewidth{\textwidth}
\tabletypesize{\footnotesize}
\tablecaption{Summary of the models considered and relative results.}
\tablehead{
\colhead{}    & \colhead{} & \colhead{} &  \multicolumn{3}{c}{$F_{diff}/F_{tot}$ (\%)} &   \colhead{}   & 
\multicolumn{3}{c}{$L_{\rm CII}$ ($10^8 \, L_{\rm \odot}$)} \vspace{1.0pt}\\
\cline{4-6} \cline{8-10} \\
\colhead{Name} & \colhead{$\langle \frac{Z}{Z_{\odot}}\rangle$} & \colhead{profile}  & \colhead{1 $\rm M_{\odot}\,yr^{-1}$}   & \colhead{10 $\rm M_{\odot}\,yr^{-1}$}    & \colhead{100 $\rm M_{\odot}\,yr^{-1}$} & \colhead{}    & \colhead{1 $\rm M_{\odot}\,yr^{-1}$}   & \colhead{10 $\rm M_{\odot}\,yr^{-1}$}    & \colhead{100 $\rm M_{\odot}\,yr^{-1}$} \vspace{1.0pt}} 
\startdata
C1 & 1& cnst& 44(\textbf{8})\phn & 37(\textbf{10})\phn & 4(\textbf{0.6})\phn &
&0.1\phn \phn&1.6 \phn \phn &21 \phd \phn \nl 
C02 & 0.2& cnst&26(\textbf{4})\phn & 20(\textbf{5})\phn & 2(\textbf{0.2})\phn &
&0.05\phn&0.7 \phn \phn &9.0 \phn \phn\nl 
C005 & 0.05 & cnst &41(\textbf{8})\phn & 20(\textbf{5})\phn & 3(\textbf{0.2})\phn &
& 0.007&0.09 \phn &1 \phn \phn\nl 
P005 & 0.05 & $\Delta$-$Z$& 24(\textbf{4})\phn & 21(\textbf{5})\phn & 2(\textbf{0.3})\phn &
& 0.05\phn&0.6 \phn \phn &8.0 \phn \phn\nl 
\enddata
\tablecomments{Name: model name, $\langle Z/Z_{\odot} \rangle$: mean metallicity in solar units, profile: type of metallicity profile adopted, $F_{diff}/F_{tot}$: percentage of the [\CII] emission arising from the CNM without (and with, in bold) the attenuation due to the increased CMB temperature, $L_{\rm CII}$: predicted [\CII] luminosity in $10^8 \, L_{\rm \odot}$.}
\end{deluxetable*}
\begin{figure}
\centering
\includegraphics[scale=0.33]{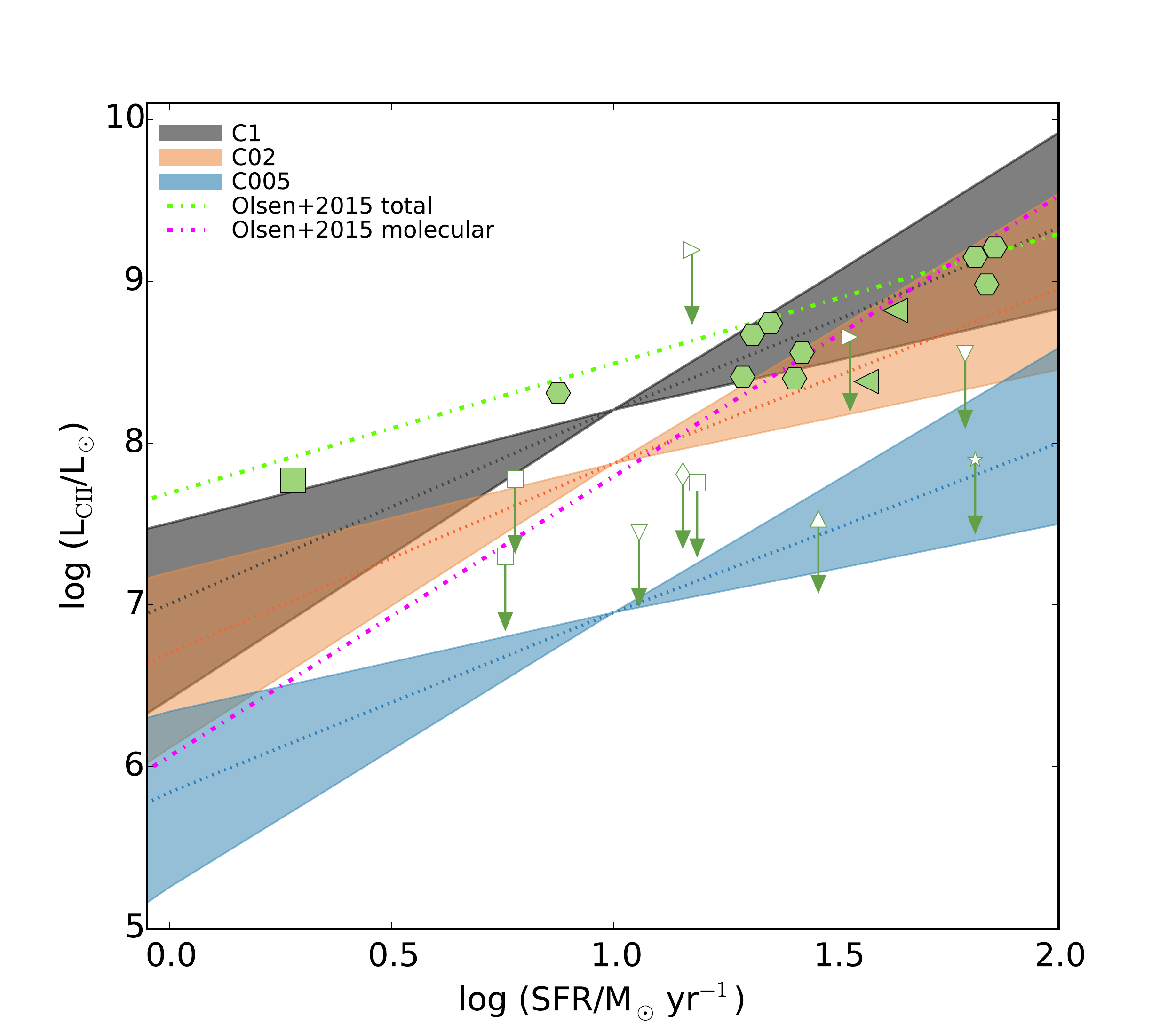}
\caption{[\CII]-SFR relation for the C1, C02, and C005 models as a function of the slope ($N$) of the KS relation, with
$N$ ranging from $0.63$ to $2.0$. We indicate with thin dotted lines the [\CII]-SFR relation obtained for our fiducial
model with ($N=1$). The [\CII] luminosity arising from all gas phases, and from molecular phase only, predicted by
\citet{olsen2015} for a sample of $z\approx2$ galaxies with metallicities ranging from $0.4\,\rm{Z_{\odot}}$ to $1.67\
\rm{Z_{\odot}}$ is plotted with green (purple) dot-dashed lines respectively. Green empty (filled) points represent upper
limits (detections) of [\CII] in LAEs and LBGs at $z\approx 5-7$, as in Fig. \ref{CII_sfr_relation}.}\label{cfr_ks_olsen}
\end{figure}
Our results contain a caveat: we have so far neglected the possible effect of stellar feedback (i.e. photo-evaporation, radiation pressure, [\HII] thermal pressure) on molecular clouds. Broadly speaking, these effects should act to reduce the mass of the molecular gas \citep{tasker2009, tasker2011, tasker2015}. On the other hand, the expansion of [\HII] regions might have either a positive effect, by triggering new star formation \citep[e.g.][]{mellema2006, bisbas2011, haworth2012},  or disperse the surrounding cloud \citep{dale2005}. Typical negative feedback timescales range from 1 to 10 Myr \citep{krumholz2006,walch2012}, namely the age of stars taken into account in our radiative transfer calculations. This would imply a steeper scaling between the SFR and molecular hydrogen surface densities, because for a given value of SFR the mass of $H_2$ is lower.

In Fig. \ref{cfr_ks_olsen} we plot the [\CII]-SFR relation for models with constant metallicity (C1, C02, and C005) as a function of the slope in the range $0.63 \leq N \leq 2.0$  \citep{shetty2013, bigiel2008, narayanan2012}. We show for reference the [\CII]-SFR relation obtained for our fiducial model $N=1$ with thin dotted lines. At a given $Z$ the steeper is the slope of the KS relation, the shallower is the [\CII]-SFR curve. Hence, a [\CII] deficit in $z\sim 6-7$ galaxies, if confirmed by deeper observations, would favor a scenario in which star formation in early galaxies blows the molecular gas apart, reducing the amount of material from which most of the [\CII] emission arises. The deviation from the local [\CII]-SFR would then imply a modified Kennicutt-Schmidt relation in $z>6$ galaxies. Stellar feedback effects are likely to be stronger in regions of very active star formation, more often located in galactic centers. This is particularly important in high-$z$ galaxies that are known to be more compact than their low-$z$ counterparts. If so, negative feedback should preferentially suppress the peak in the [\CII] spectrum at the systemic redshift of the galaxy. Alternatively/in addition, as can be noted in Fig. \ref{cfr_ks_olsen}, the deficit might be explained by lower gas metallicities.
In the same Figure, we also test our results with the [\CII]-SFR relation for all gas phases (PDR + molecular + ionized) and from molecular phase only, calculated by \citet{olsen2015} in a sample of simulated $z\approx2$ galaxies. As they point out in their work, their PDR component is best identified with what we call CNM, while their molecular component, located at the center, is comparable to what we call PDR emission.
We find a nice agreement between our results and their findings for the molecular gas. This is something expected, given that the fraction of [\CII] emission arising from the CNM is almost totally attenuated due to the increased CMB temperature at $z \approx 6.6$.

Finally, we note that the MC density distribution may play a role. Our simulated galaxy is characterized by a mean molecular hydrogen number density $n_{\rm cl}\sim 10^{2.9}$cm$^{-3}$; this quantity depends on the square of the assumed Mach number $\mathcal{M} =10$. Calculations performed with \texttt{UCL\_PDR} show that molecular clouds characterized by densities 10 times higher (lower), for a fixed gas metallicity (e.g. $ {\rm log}(Z/Z_{\odot})=-1.5$), would result into a [\CII] emissivity 5 times higher (20 times lower) than found here. Although we consider such large variations of the Mach number unlikely, at present we cannot exclude that the corresponding shift in the mean MC density plays some role in the interpretation of the results.

\bibliography{CII_SFR_relation}
\end{document}